\documentclass[twocolumn,twoside,slac_two]{revtex4}
\usepackage[utf8]{inputenc}
\usepackage{graphicx}
\usepackage{fancyhdr}
\usepackage{graphics}
\usepackage{epstopdf}
\usepackage{textpos}
\begin{document}

%%%%%%%%%%%%%%%%%%%%%% WRITE THE TITLE HERE %%%%%%%%%%%%%%%%%%%
\title{\centering Performance of the ATLAS tau trigger during 2011 data taking period}
%%%%%%%%%%%%%%%%%%%%%% WRITE THE AUTHOR HERE %%%%%%%%%%%%%%%%%

\author{
\centering
\begin{center}
Charlie Isaksson on behalf of the ATLAS Collaboration
\end{center}}
\affiliation{\centering Department of Physics and Astronomy, Uppsala University, Lägerhyddsvägen 1, 752 37 Uppsala, Sweden}
%%%%%%%%%%%%%%%%%%%%%% WRITE THE ABSTRACT HERE %%%%%%%%%%%%%%%%
\begin{abstract}
Many models for physics Beyond the Standard Model predict an increased production rate of tau leptons. Therefore hadronically de\-caying tau leptons play an important role in searches for new physics at the LHC. By triggering on them, the discovery power of ATLAS can be greatly enhanced for many Beyond the Standard Model searches. In this contribution we present a brief description of the ATLAS tau trigger system and demonstrate its robustness and reliability during 2011.
\end{abstract}

%%%%%%%%%%%%%%%%%%%%%%%%%%%%%%%%%%%%%%%%%%%%%%%%%%%%%%%%%%
\maketitle
\thispagestyle{fancy}

\section{Introduction}
The ATLAS detector at the LHC has been designed to search for any hint
of new physics, such as the Higgs boson, Supersymmetry or exotic heavy resonances. 
The tau lepton plays an important role in many of these searches and triggering on
hadronically decaying\linebreak taus increases the sensitivity to new physics~\cite{CSCbook}.

\section{The ATLAS tau trigger}
Hadronic decays of tau leptons consist of mainly one or three charged pions, possible neutral pions and a tau neutrino.
They leave a jet-like signature in the detector and can be distinguished from QCD jets by their isolation, low track multiplicity
and the energy deposition in a narrow cone in the calorimeter. These characteristics of tau leptons are exploited by the tau trigger.

\subsection{Level 1}
The L1 tau trigger is a simple and fast hardware trigger that uses ElectroMagnetic (EM) and hadronic calorimetry trigger towers of size $\Delta \eta \times \Delta \phi = 0.1 \times 0.1$. At this level tau lepton candidates are identified using\linebreak three key features, the energy in two adjacent EM towers, the energy in $2 \times 2$ pairs of hadronic towers\linebreak behind the EM cluster and the EM and hadronic\linebreak energy in an isolation region between the $2 \times 2$ core and a $4 \times 4$ ring. A given L1 threshold will set a\linebreak requirement on the previous quantities. Current settings apply a lower threshold at L1 and use the identi\-fication power and better energy determination at subsequent levels.

\subsection{Level 2}
At L2 more complex selections are applied, using\linebreak refined calorimeter lateral shape and transverse\linebreak energy variables. Tracks are also reconstructed in\linebreak regions passing the L1 trigger using the full detector gra\-nularity. The characteristic track and calorimeter\linebreak narrowness and low track multiplicity of the tau\linebreak lepton is used to discriminate against background. 

\subsection{Event Filter}
At the Event Filter (EF) level, parts of the offline tau reconstruction algorithms are used on the seeds passing L2. Data from the whole detector can be\linebreak accessed if necessary. This provides a wide range of more accurate identification variables. 

\begin{figure}
\includegraphics[width=0.9\linewidth]{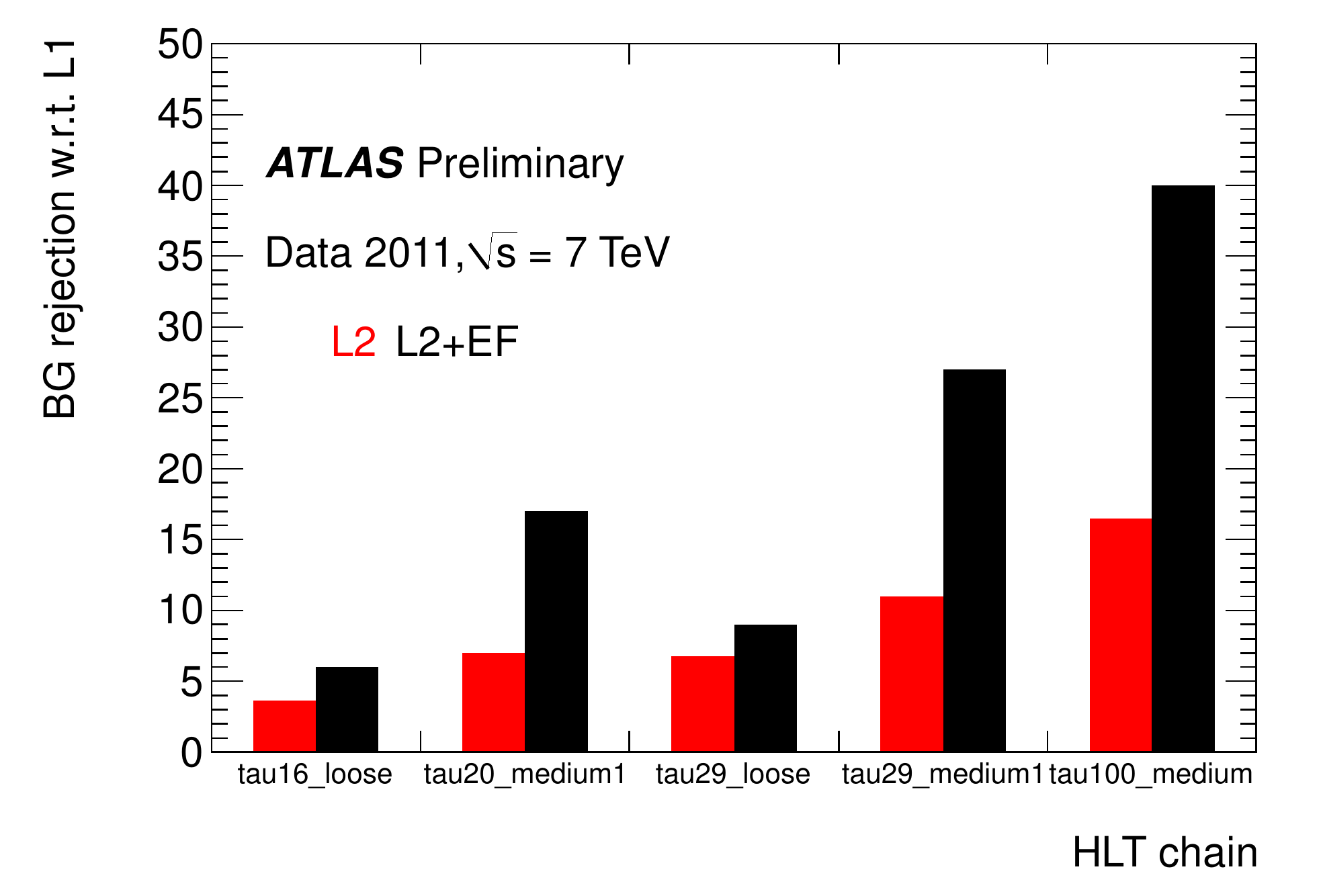}
\caption{\label{jetrej} QCD jet rejection factors of different High Level Trigger tau chains determined from collision data taken in 2011. The numbers given are with respect to the output of the associated L1 tau trigger item. The red and black bars show the rejection after L2 and EF, respectively~\cite{pubtwiki1}.}
\end{figure}

Rejection against background by the High Level Triggers (L2 together with EF and referred to as HLT) is of the order of 10 or more, depending on $p_T$ range and tightness in selection (see Fig.~\ref{jetrej}).

\subsection{Trigger Menu}
Different tau trigger signatures have been used for early running and for increasing instantaneous luminosities. Typically, the $p_T$ threshold applied at EF is tightened with increasing luminosities. Additionally, different quality requirements (loose, medium, tight) are available for each chain. Chains with a loose qua\-lity requirement currently only appear in combined chains, but were used for single tau triggers at lower luminosities. 

A subset of the ATLAS tau triggers and their pur\-pose is shown in Table~\ref{trigmen}.

\begin{table}[t]
\caption{\label{trigmen}A sample of ATLAS tau triggers.}
\begin{tabular}{|c|c|c|}
\hline \textbf{Menu} & \textbf{Purpose} & \textbf{High Level Trigger}\\
\hline Single Tau & $H^+$, SUSY & tau100\_medium \\ 
\hline 2 Tau & H,Z & 2tau29\_medium \\ 
\hline Tau+Lepton & H,$t\bar{t}$,Z & tau16\_loose\_mu15 \\
&                & tau16\_loose\_e15\_medium \\ 
\hline Tau+MET    & $H^+$,$W \rightarrow \tau\nu$ & tau29\_medium\_xe35\_noMu \\
&                               & tau29\_loose\_xs70\_loose\_noMu \\
&                               & tau29\_loose1\_xs45\_noMu\_3L1J10 \\
\hline
\end{tabular}
\end{table}

\section{Performance}
\subsection{Trigger Rates}
\begin{figure}
\includegraphics[width=0.9\linewidth]{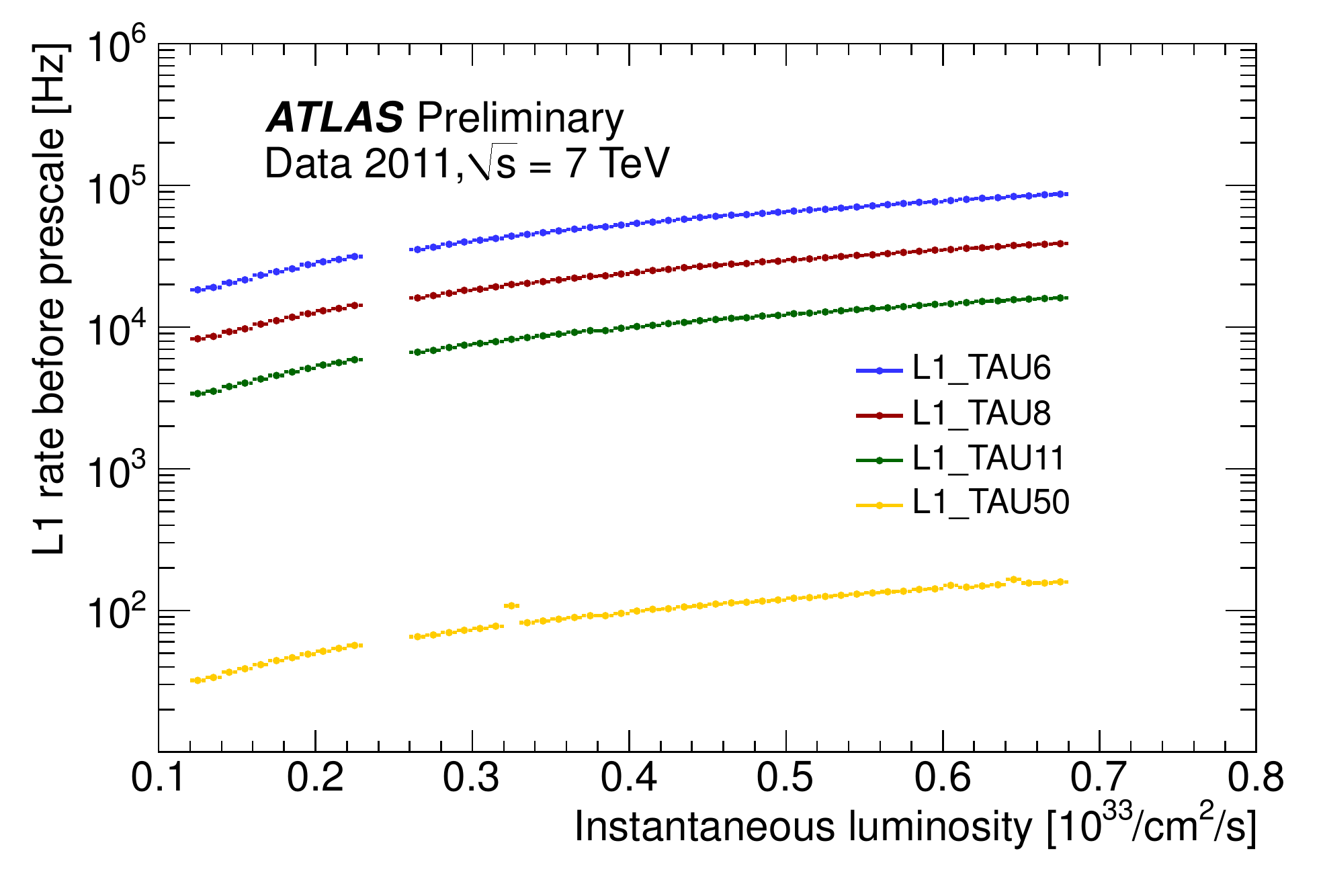}
\caption{\label{l1rates}L1 rates before prescale versus the instantaneous luminosity measured by ATLAS for four different L1 tau items that are feeding primary HLT tau chains. The last digits in the L1 item name correspond to the $E_T$ requirement, e.g. a L1 $E_T >$ 11 GeV for L1\_TAU11~\cite{pubtwiki1}.} 
\end{figure}

\begin{figure}
\includegraphics[width=0.9\linewidth]{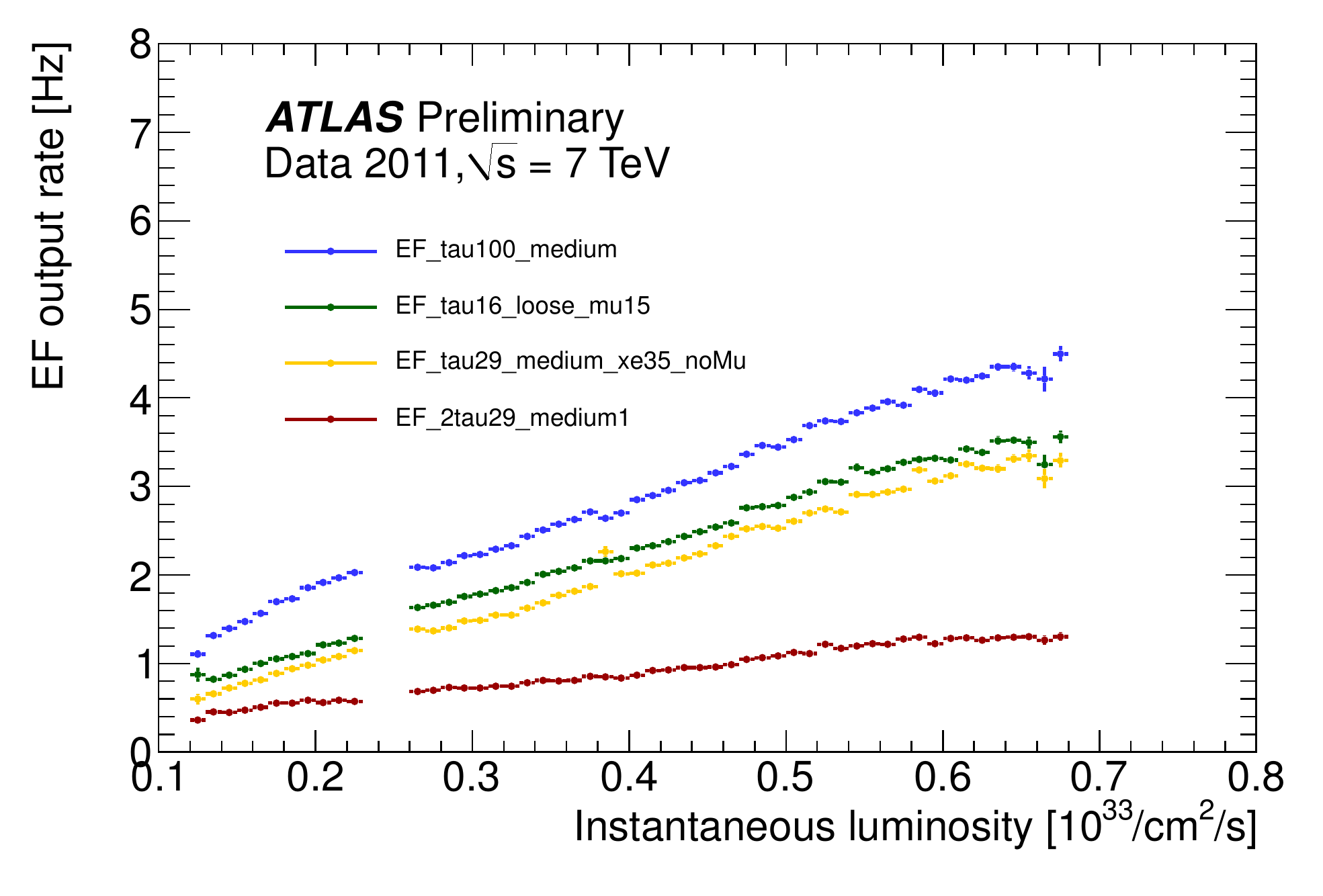}
\caption{\label{efrates}EF output rates versus the instantaneous luminosity measured by ATLAS for four selected HLT tau chains. The numbers in the item names correspond to $E_T$ requirements at EF for a given trigger type, e.g. a muon candidate with $E_T >$ 15 GeV in mu15 or missing transverse energy $>$ 35 GeV in xe35~\cite{pubtwiki1}.}
\end{figure}

The primary goal of the ATLAS triggers is to reduce\linebreak the rate of events to a manageable level for disk sto\-rage by selecting events with interesting characte\-ristics and discarding the several orders of magnitude larger background of uninteresting events. While the maximum speed at which data can be written to disk is only about 200 Hz, the rate before L1 is typically around 40 MHz (depending on luminosity) and about 75 kHz after the faster ($\sim2.5\mu$s) L1 trigger. However with the greater identification power of the reconstruction algorithms used in the HLT the rates are brought down to $\sim5$ kHz after L2 and below $\sim10$ Hz after the EF trigger. Fig.~\ref{l1rates} and Fig.~\ref{efrates} show the rates of different trigger chains for L1 and EF respectively. 

\begin{figure}
\includegraphics[width=0.9\linewidth]{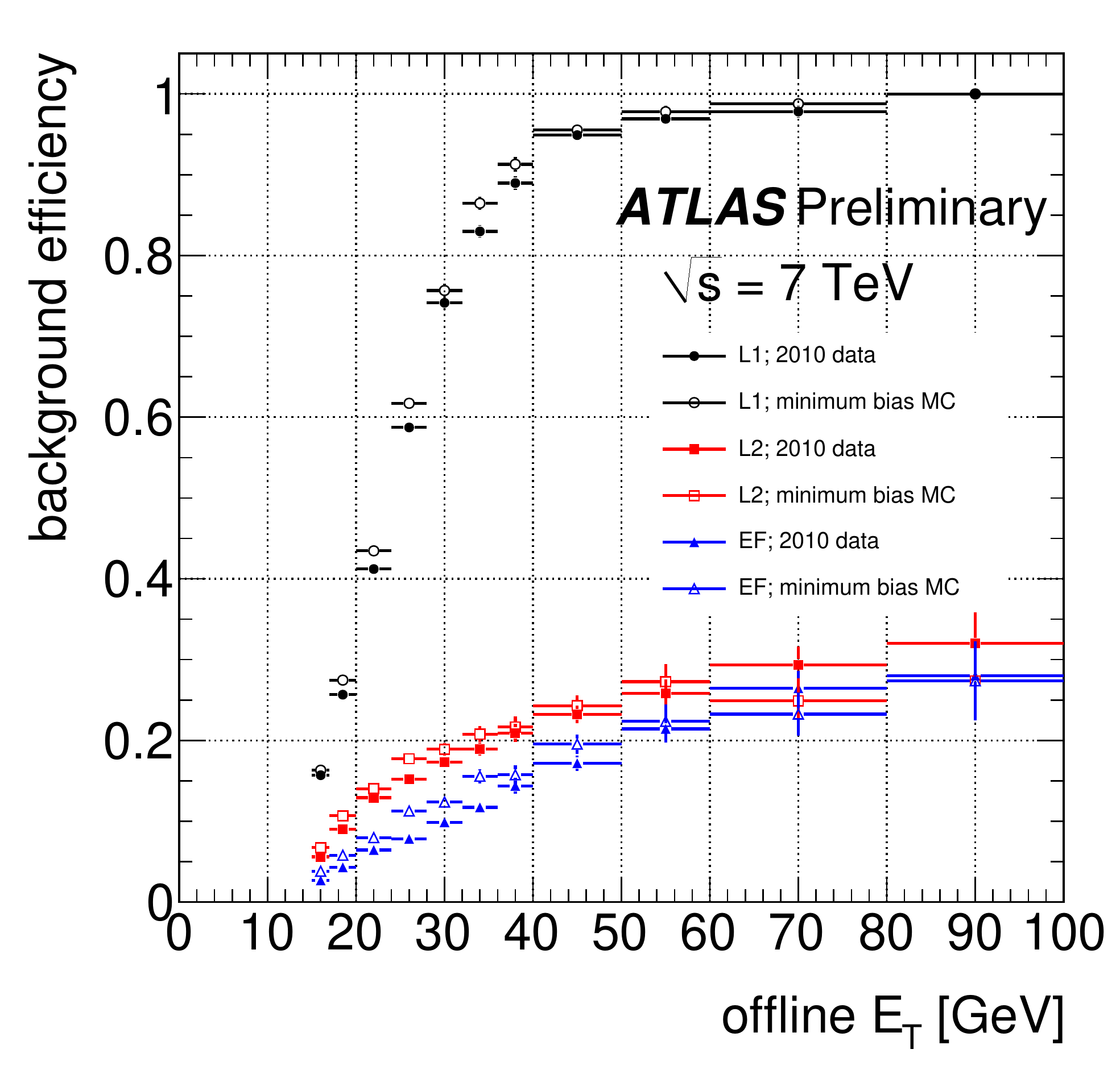}
\caption{\label{bgeff}Fraction of the background offline tau candidates passing L1, L2 and EF tau12\_loose trigger as a function of the $E_T$ of the offline tau candidate. Requirements for $E_T$ are 5 GeV at L1, 7 GeV at L2 and 12 GeV at EF. Distributions are done on Monte-Carlo (MC) Minimum Bias simulated with PYTHIA and on data from Minimum Bias stream~\cite{pubtwiki2}.}
\end{figure}

\subsection{Trigger Efficiency}
While keeping the event rates low it is also important for the tau trigger to successfully distinguish between real tau leptons and backgrounds in which a tau signature is faked by a QCD jet or an electron. The ATLAS tau trigger has been successful in keeping the trigger efficiency low for background as is shown in Fig.~\ref{bgeff}. Without correcting for underlying event, the simulation reproduces well the observed background rejec\-tion performance in the data. Tuned simulations are expected to show better agreement with data results.

\begin{figure}
\includegraphics[width=0.9\linewidth]{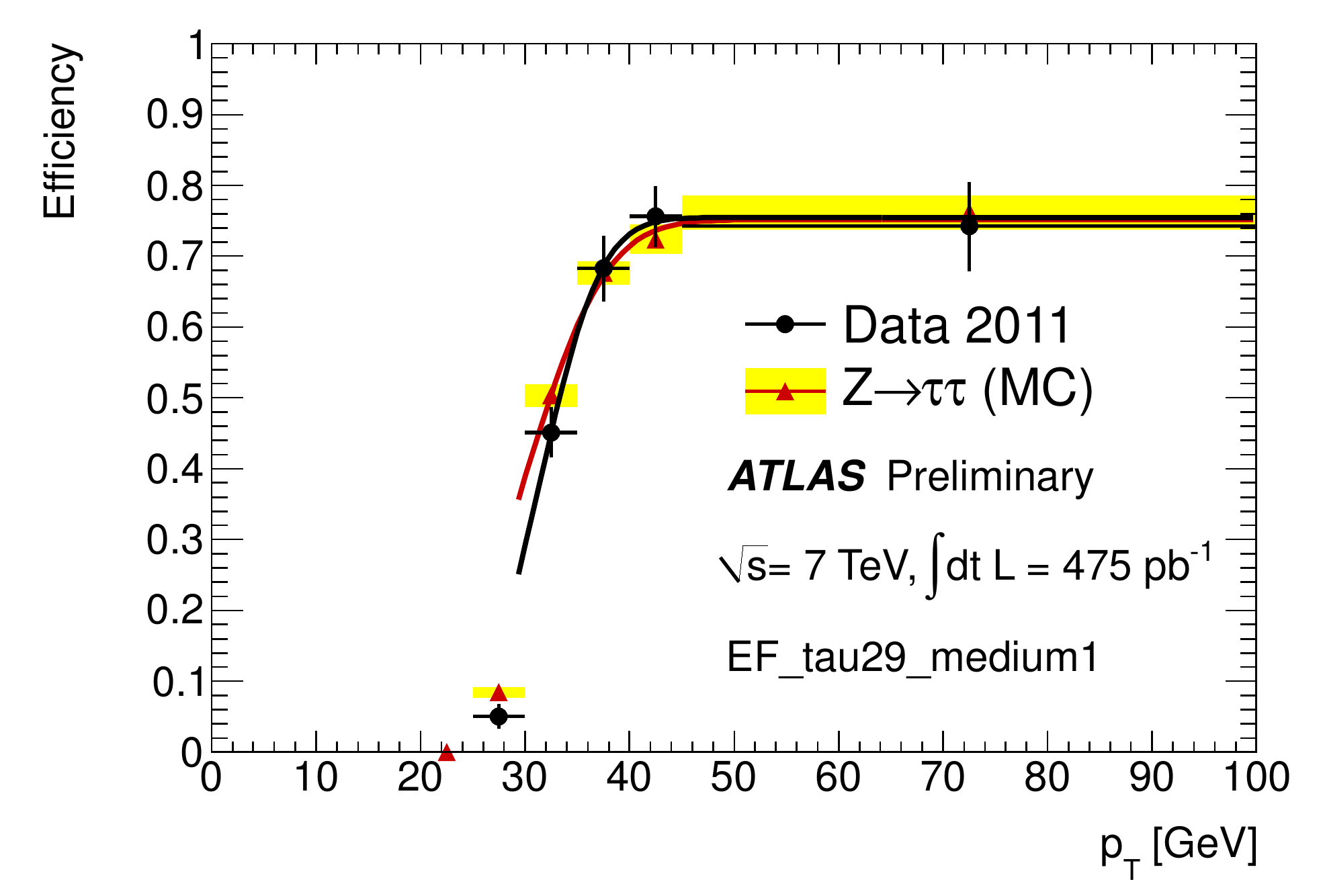}
\caption{\label{sigeff}Efficiency of the EF\_tau29\_medium1 trigger chain with respect to offline reconstructed tau candidates, as a function of the offline $p_T$. The trigger includes a strict requirement on the number of tracks associated to the trigger object in addition to the regular 'medium' selection. The measurement was made using a tag and probe analysis with $Z\rightarrow\tau\tau\rightarrow\mu h$ events in 2011 data~\cite{pubtwiki3}. The tau candidates are required to pass medium identification criteria. The analysis follows closely the method from the $Z\rightarrow\tau\tau$ cross-section measurement~\cite{zttcross}.}
\end{figure}

The tau trigger has also been successful in keeping the efficiency high for real taus. The trigger  efficiencies have been studied in $Z\rightarrow\tau\tau\rightarrow\mu h$ events selec\-ted with a tag-and-probe technique in 2011 data. The observed efficiency is shown in Fig.~\ref{sigeff}.

\section{Conclusions}
ATLAS has run very successfully during 2011,\linebreak collecting more than $5fb^{-1}$ of data at $\sqrt{s} =$ 7~TeV. The tau trigger has performed remarkably well, and was successfully used in the first observation of\linebreak $W \rightarrow \tau\nu$ events by ATLAS in 2010~\cite{firstW}, and in 2011 in the measurement of the $W \rightarrow \tau\nu$ cross section for instance~\cite{wxs}. Its great performance so far gives us confidence about its accuracy, and we are sure it will\linebreak remain a vital asset in searching for new physics among the vast amount of data collected.

\vfill

\end{document}